\documentclass[12pt]{article}
\usepackage{epsf}
\begin{document}
\baselineskip 18pt
\begin{center}
{\Large Further Considerations on the CP Asymmetry in}\\
{\Large Heavy Majorana Neutrino Decays}
\vskip .75in
{\large Marion Flanz, Emmanuel A. Paschos}
 
{\it Institut f\"{u}r Physik},\\ {\it Universit\"{a}t Dortmund},\\
{\it 44221 Dortmund, Germany}\\

\end{center}

\vskip .75in
\begin{abstract}
\baselineskip 16pt

We work out the thermodynamic equations for the decays and scatterings 
of heavy Majorana neutrinos including the constraints from unitarity.
The Boltzmann equations depend on the CP asymmetry parameter which
contains both, a self-energy and a vertex correction. At thermal
equilibrium there is no net lepton asymmetry due to the 
CPT theorem and the unitarity constraint. We show explicitly that deviations 
from thermal equilibrium create the lepton asymmetry. 
\end{abstract}

\newpage
\section{Introduction}
\baselineskip 18pt

%There are many possibilities to explain the Baryon asymmetry of the 
%universe (BAU). One of them is the extension of the Standard model (SM)
%with a heavy Majorana neutrino (HMN) for each generation. The masses of
%the HMN are above the electroweak breaking scale and below any inflation
%or Grand Unification Theory (GUT) scale.

%At the beginning of the expansion of the universe, the universe was very
%hot and the HMN are in thermal equilibrium. They decay and recombine and
%scatter at each other and at the other particles. When the universe cools
%down the temperature fell below the mass of one of the HMN and from now
%on this particle can only decay, but will not be created through a 
%recombination from other particles, because their energy will not suffice.  
%
%The development of the particle density with the expansion of the 
%universe is given through the Boltzmann equations (BE). 
 
Over the past few years it has been shown that a lepton asymmetry can
be generated by the mixing of heavy Majorana neutrinos. Majorana neutrinos
have the remarkable property of being mixed states of particles and 
antiparticles. By definition they induce $\Delta L = 2$ and
consequently $\Delta (B-L) = -2$
transitions. In addition they could have couplings to scalar particles,
which allow them to 
decay into Higgs particles and leptons, i.e. $ N \rightarrow
\phi^\dagger \ell, \phi \, \ell^c$ \cite{fuku}. 
In these models $CP-$violation is introduced through complex
couplings and appears in the self-energy \cite{flanz, flanzjan}
or vertex corrections \cite{fuku, luty, buch, mich}. We shall
classify the effects using the terminology of $K^0$ mesons \cite{flanzjan}. 
We shall
call direct or $\varepsilon^\prime-$type effects, the $CP-$violation 
which arises from the interference of tree diagrams with 
vertex corrections. Similarly we call indirect or $\delta-$type effects,
those which arise through the self-energies. The self-energies play
the role of the box diagram in the $K^0$ system and define the
physical states.
A consequence of the observation in
\cite{flanz, flanzjan} is that the physical Majorana states $\Psi$
are not $CP$ eigenstates and that the $CP-$asymmetries 
produce observable effects. Finally, 
the mass splitting between the states also plays a role 
and when the mass difference is of the order of the
width there is a resonance enhancement \cite{flanzjan}.

The processes under discussion must satisfy unitarity constraints
so that the sum of the probabilities for all transitions to and from
a state $i$ should sum to one and yields \cite{dolg, wolf}:
\begin{equation}
\sum_j |{\cal M}(i \rightarrow j)|^2 = \sum_j |{\cal M}(j \rightarrow i)|^2
\quad . \label{uni} 
\end{equation}
For our case this means that when we consider the scatterings 
$ \ell \phi^\dagger \leftrightarrow \ell^c \phi $ equation (\ref{uni})
turns into
\begin{eqnarray}
\lefteqn{|{\cal M}(\ell \phi^\dagger \rightarrow \ell \phi^\dagger)|^2 +
|{\cal M}(\ell \phi^\dagger \rightarrow \ell^c \phi)|^2} & & \label{wei} \\
 & = & |{\cal M}(\ell \phi^\dagger \rightarrow \ell \phi^\dagger)|^2 +
|{\cal M}(\ell^c \phi \rightarrow \ell \phi^\dagger)|^2 \nonumber
\quad .  
\end{eqnarray}
so that the probabilities for the direct and inverse process should be
equal.
For the early universe a summation over all possible physical states
with Boltzmann factors weighting the inital states is required.
Boltzmann factors are introduced according to the
thermal properties of the universe. At thermal equilibrium, for
example, the weighting factors are all equal, which
implies that the lepton asymmetry averages to zero \cite{dolg}. 
This issue and its implications are 
explicitly proved in \cite{franc} and further discussed in \cite{plue}.

The generation of a lepton or baryon asymmetry is a combination of the 
Sakharov effects \cite{sak} (i) baryon or lepton violation, 
(ii) $C$ and $CP-$violation and (iii) the thermodynamic properties
of the ensemble of particles in the early universe. 
For this reason 
several early papers studied the non-equilibrium equations 
of the ensemble \cite{wolf, adol, kolb}. These thermal considerations 
together with several subtleties of the phenomenon are 
mentioned in \cite{luty}. In this article we present an 
explicit calculation of the modification of the unitarity constraints 
as implied by
the thermodynamics of the system and point out the meaning of the
various terms. In fact we shall show that 
the s-channel physical poles play a special role and are responsible for the
development of asymmetries.

To begin with we consider {\em massive unstable particles} which
interact and decay. As the particles interact
some of them occur in intermediate states. When the lifetimes of 
the particles are long relative to the interaction times, the
intermediate states interact many times with the thermal bath.
For this reason they should not be considered as virtual states, but
as ensembles of particles with their own thermodynamic distributions.
This phenomenon modifies the unitarity sum, i.e. the various terms
in equation (\ref{wei}) are weighted by different factors, and a lepton
asymmetry is created.

We estimate the collision of the particles in relation to their 
lifetimes. The width of the particles with energy $E$ and mass $M_N$
is given by
\begin{equation}
\Gamma_N = \frac{M_N}{E} \, 
\Gamma_0 = \frac{|h_\ell|^2}{16 \pi} \frac{M_N^2}{T}
\quad \quad {\rm for} \quad E = T \quad .
\end{equation}
Their frequent interactions are like $ e^- + t \rightarrow N + b$
with a Higgs particle exchanged in the t-channel. The cross-section 
at temperatures $T = E \gg M_N$ is
\begin{equation}
\sigma = \frac{|h_t|^2 |h_\ell|^2}{16 \pi} \frac{1}{E^2} \quad .
\end{equation} 
We now calculate the interaction of one particle $N$ with a density
of quarks. At a temperature $T \gg M_N$ and assuming Boltzmann statistics
the particle density is given by
\begin{equation}
n = \frac{2}{\pi^2} \, T^3 \quad .
\end{equation} 
Therefore the interaction rate for relativistic velocities is 
\begin{equation}
n \cdot \sigma \cdot v = \frac{|h_t|^2 |h_\ell|^2}{8 \pi^3} \, T \quad .
\end{equation}
Thus there are many more interactions per decay width
\begin{equation}
\frac{n \cdot \sigma \cdot v}{\Gamma_N} \sim O\left( \left( \frac{T}{M_N} \right)^2
|h_t|^2 \right) \quad ,
\end{equation}
with the last factor being of the order of $10^6$ or larger. Since
many scatterings are taking place during the lifetime of the particles,
each state has a different history and the particle states are incoherent.
At this epoch when we consider many scatterings, decays and recombinations
of particles, there is no asymmetry, because the unitarity of the S-matrix
guarantees the equality of direct and inverse processes. An asymmetry
is generated when specific channels begin to decouple.

As usual we work on an extension of the standard model where we 
include one heavy right-handed Majorana field $N_{i}$ per generation of
light-lepton $(i = 1, 2, 3).$ The new fields are singlets with
respect to the standard model \cite{fuku}. Extensions of this model
have been constructed in gauge theories \cite{apost, apostel},
supersymmetric theories \cite{covi, mipl} and theories with electroweak
singlet neutrinos \cite{akh}.

\begin{figure}[h]
\begin{center}
\mbox{\epsfbox{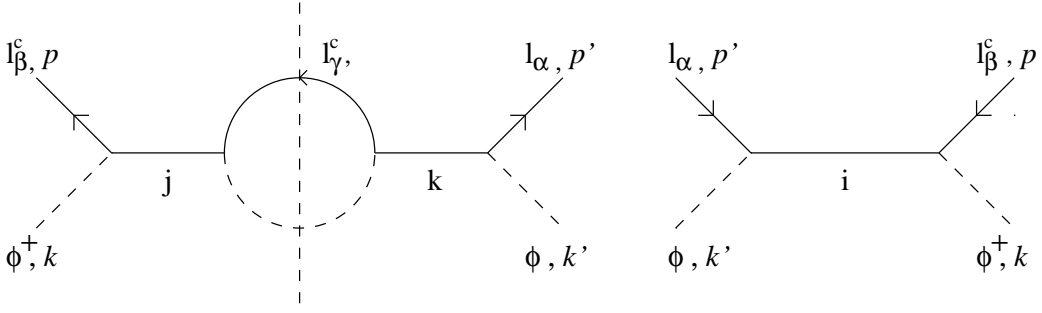}}
\caption{Interference of the tree with the one-loop self-energy diagram}
\end{center}
\end{figure}

We consider the contribution from the interference of the diagrams
in Figure 1,
which appears in the asymmetry of the reaction $ \ell^c \phi 
\rightarrow \ell \phi^\dagger.$ In the figure,
we have indicated the momenta
explicitly and have taken the absorptive part of the self-energy.
In addition we consider many generations indicated by the indices 
$\alpha$ and $\beta$ for the external particles, the index $\gamma$ for
the loop-momentum and $i, j$ and $k$ for the heavy particle propagators.
The contribution to the asymmetry is
\begin{eqnarray}
2 {\rm Re}({\cal M}_L^\ast {\cal M}) & = & 
 \frac{E E^\prime}{2\pi^2} (kp + pp^\prime - kp^\prime)
 \frac{M_i M_k \, {\rm Im}(h_{\beta i}^\ast
 h_{\alpha i}^\ast h_{\alpha k} h_{\gamma k} h_{\gamma j}^\ast
 h_{\beta j})
  }{(s - M_i^2)(s - M_k^2)(s - M_j^2)} \nonumber \\
 & = &  \frac{E E^\prime}{2\pi^2} (kp + pp^\prime - kp^\prime)
 \frac{M_i M_k \, {\rm Im}(h_{\gamma k} h_{\alpha k}
 h_{\alpha i}^\ast h_{\beta i}^\ast  h_{\beta j}  
h_{\gamma j}^\ast)}{(s - M_i^2)(s - M_k^2)(s - M_j^2)}
\end{eqnarray}
with ${\cal M}_L$ is the loop diagram and ${\cal M}$ is the Born diagram
amplitude (see Figure 1).
Carrying out the same calculation for the process $\ell \phi^\dagger
\rightarrow \ell^c \phi$ we obtain the following expression for the
asymmetry:
\begin{equation}
2 {\rm Re}(\overline{{\cal M}_L^\ast} \, \overline{{\cal M}}) = 
\frac{E E^\prime}{2\pi^2} (kp + pp^\prime - kp^\prime)
 \frac{M_i M_k \, {\rm Im}(h_{\beta i}
 h_{\alpha i} h_{\alpha k}^\ast h_{\gamma k}^\ast h_{\gamma j}
 h_{\beta j}^\ast)
}{(s - M_i^2)(s - M_k^2)(s - M_j^2)} \quad .
\end{equation}
In this expression we kept the propagator field for all intermediate
states but omitted their widths. The two expressions are the same 
except for the factor of the coupling constants. 
A careful comparison of the two expressions
shows that they become identical when we sum over the intermediate
states and over all external particles. This can be seen by 
rearranging of the indices. It is a consequence of
unitarity mentioned above. The same property persists when we keep
all the masses of the propagators or when we replace the propagators
with $\delta-$functions in the narrow-width approximation. The 
two expressions $2 {\rm Re}({\cal M}_L^\ast {\cal M})$ and
$2 {\rm Re}(\overline{{\cal M}_L^\ast} \, \overline{{\cal M}})$ 
are different when we
select specific components. This is in fact what happens in the 
development of the universe, as the temperature reaches the mass 
of a heavy neutrino specific components are multiplied by special
Boltzmann factors, which single them out as demonstrated below.

\section{The Boltzmann Equations}
In this section we explicitly discuss the thermodynamic development
of the system. In this way we explain how deviations from thermal
equilibrium modify the sum of unitary contributions and render them
different from zero. The lepton asymmetry is a sum of several terms:

\noindent
First, there are the decays of the heavy Majorana neutrinos:
\begin{itemize} 
\item[] $ \Psi_i \rightarrow \ell \phi^\dagger$
which creates an excess of leptons, and
\item[] $ \Psi_i \rightarrow \ell^c  \phi$ which reduces the 
amount of leptons 
\end{itemize}
and the corresponding recombination terms. In addition there
are two scattering processes:
\begin{itemize} 
\item[] $\ell^c \phi \rightarrow \ell \phi^\dagger$
which creates an excess of leptons, and
\item[] $\ell \phi^\dagger \rightarrow \ell^c  \phi$ which reduces the 
amount of leptons. 
\end{itemize}
The result of all these terms is the following differential equation
for the lepton asymmetry density $n_L = n_\ell - n_{\ell^c}:$ 
\begin{eqnarray}
\dot{n}_L + 3 H n_L & = & 
 f_{\Psi_1} \left[ |{\cal M}(\Psi_1 \rightarrow 
          \ell \phi^\dagger)|^2  - |{\cal M}(\Psi_1 \rightarrow 
          \ell^c \phi)|^2 \right] \Lambda^{3}_{12} \label{bolle} \\
 & & + \left[ 
    - f_\ell f_{\phi^\dagger} |{\cal M}(\ell \phi^\dagger \rightarrow
    \Psi_1)|^2 + f_{\ell^c} f_\phi |{\cal M}(\ell^c 
     \phi \rightarrow \Psi_1)|^2 \right] \Lambda_{12}^{3} \nonumber \\  
 & & + 2 \, \Lambda_{12}^{34} \left\{ f_{\ell^c} f_\phi \left[
     |{\cal M}(\ell^c \phi \rightarrow \ell \phi^\dagger)|^2 -
     |{\cal M}_{RIS}(\ell^c \phi 
      \rightarrow \ell \phi^\dagger)|^2 \right] \right. \nonumber \\
  & & - \left. f_\ell f_{\phi^\dagger} \left[
     |{\cal M}(\ell \phi^\dagger \rightarrow \ell^c \phi)|^2 -
     |{\cal M}_{RIS}(\ell \phi^\dagger 
      \rightarrow \ell^c \phi)|^2 \right] \right\} \nonumber
\end{eqnarray}
where $\Lambda_{12}^{34}$ is the four-particle phase space factor 
depending on the momenta $p_1$ to $p_4,$  
$f_\ell$ and $f_{\phi^\dagger}$ are the Boltzmann distributions for
leptons and scalar particles, respectively. 
$f_\Psi$ is the distribution for the
heavy Majorana states. Since the $\Psi_1'$s are produced copiously 
through the multiple scatterings as discussed in the previous section
they have their own distribution. 
For the decays and recombinations we introduce the three-particle phase
space $\Lambda_{12}^3.$
The first two lines already include the decays of the real intermediate 
states (RIS) and their recombinations. For this reason they are
subtracted from the scattering amplitudes.

The interference of the tree with the one loop graph gives the
$CP-$\-vio\-la\-ting factor:
\begin{equation}
\left[ |{\cal M}(\Psi_1 \rightarrow 
          \ell \phi^\dagger)|^2  - |{\cal M}(\Psi_1 \rightarrow 
          \ell^c \phi)|^2 \right] =
(\varepsilon^\prime + \delta) \, |{\cal M}_0|^2 
\end{equation}
with $|{\cal M}_0|^2$ being the tree level amplitude.
Both $\varepsilon^\prime$ and $\delta$ are $CP$ violating 
parameters with $\varepsilon^\prime$ is produced by the vertices
(direct $CP-$violation) and $\delta$ from the self energies.
Combining the terms and neglecting those ones of $O(\left[
\varepsilon^\prime + \delta \right]^2, 
\left[ \mu (\varepsilon^\prime + \delta) \right]/T),$ 
the first two lines of equation (\ref{bolle}) give:
\begin{equation}
n_{\Psi_1} \, (\varepsilon^\prime + \delta) \, \langle \Gamma_{\Psi_1} 
\rangle + n^{EQ}_{\Psi_1} \, (\varepsilon^\prime + \delta) \, 
\langle \Gamma_{\Psi_1} \rangle - \left( \frac{\mu}{T} \right) 
n^{EQ}_{\Psi_1} \, 
\langle \Gamma_{\Psi_1} \rangle \quad , \label{wild}
\end{equation}
where the phase space was combined with the amplitude $|{\cal M}_0|^2$
to produce the thermally-averaged decay width 
$\langle \Gamma_{\Psi_1} \rangle.$

Next we consider the last two lines of equation (\ref{bolle}), where
we expand the particle distributions in powers of $(\mu/T):$
\begin{eqnarray}
\lefteqn{ }
 & & 2 \, \Lambda_{12}^{34} \left[ f_{\ell^c} f_{\phi}
 |{\cal M}(\ell^c \phi \rightarrow \ell \phi^\dagger)|^2 -
 f_{\ell} f_{\phi^\dagger}
 |{\cal M}(\ell \phi^\dagger \rightarrow \ell^c \phi)|^2 \right.
 \nonumber \\
 & & \left. + f_\ell f_{\phi^\dagger} 
  |{\cal M}_{RIS}(\ell \phi^\dagger \rightarrow \ell^c \phi)|^2 -
  f_{\ell^c} f_\phi 
  |{\cal M}_{RIS}(\ell^c \phi \rightarrow \ell \phi^\dagger)|^2 
 \right] \label{firle} \\
& = & 2 \, \Lambda_{12}^{34} \, f^{EQ}_{\Psi_1} \left[
 (1 - \frac{\mu}{T}) 
 |{\cal M}(\ell^c \phi \rightarrow \ell \phi^\dagger)|^2 -
  (1 + \frac{\mu}{T}) 
 |{\cal M}(\ell \phi^\dagger \rightarrow \ell^c \phi)|^2 \right.  
 \nonumber \\
& & + \left.
 (1 + \frac{\mu}{T}) |{\cal M}_{RIS}(\ell \phi^\dagger \rightarrow 
 \ell^c \phi)|^2  -
  (1 - \frac{\mu}{T}) 
|{\cal M}_{RIS}(\ell^c \phi \rightarrow \ell \phi^\dagger)|^2 
 \right] \nonumber \\
 & & + O\left( \frac{\mu^2}{T^2} \right) \nonumber 
\end{eqnarray}
The leading terms proportional to unity involve the difference
of the complete amplitudes $|{\cal M}|^2$ and vanish by virtue of
{\it unitarity}. This is the only place where unitarity is
effective in bringing a complete
cancellation. The remaining terms arise from interactions
which change the number of particles (chemical potentials) and through
deviations from equilibrium. The former is the origin of terms proportional
to the chemical potentials which are multiplied with $CP-$ conserving
amplitudes. The leading term from the real intermediate
states is of special interest, because it is twice as big as the
second term in equation (\ref{wild}), but has the opposite sign.
Its net effect is to change the sign of the second term in
equation (\ref{wild}). After some algebra, described in the appendix, 
equation (\ref{firle}) leads to:
\begin{eqnarray}
 & & - 2 \, n^{EQ}_{\Psi_1} (\varepsilon^\prime + \delta)  
 \langle \Gamma_{\Psi_1} \rangle \label{kai} \\
 & &
- 2 \, \left( \frac{\mu}{T} \right) n_\gamma^2 
\left[ \langle \sigma^\prime(\ell \phi^\dagger \rightarrow 
 \ell^c \phi) \cdot v \rangle + \langle 
 \sigma^\prime(\ell^c \phi \rightarrow \ell 
\phi^\dagger) \cdot v \rangle \right] \nonumber \quad .
\end{eqnarray}
We can substitute the chemical potential by the density for the 
lepton asymmetry using the equation
\begin{equation}
\frac{n_L}{n_\gamma} = 2 \, \frac{\mu}{T} + O\left( \frac{\mu^3}{T^3}
  \right) \quad .
\end{equation}
Combining the various terms we arrive at the 
final Boltzmann equation:
\begin{eqnarray}
\dot{n}_L + 3 H n_L & = & (\varepsilon^\prime + \delta) \left[
n_{\Psi_1} - n_{\Psi_1}^{EQ} \right] \langle \Gamma_{\Psi_1} \rangle  
- \frac{1}{2} \left( \frac{n_L}{n_\gamma} \right) n^{EQ}_{\Psi_1} 
  \langle \Gamma_{\Psi_1} \rangle \label{abi} \\
 & & - n_L \, n_\gamma 
\left[ \langle \sigma^\prime(\ell \phi^\dagger \rightarrow 
 \ell^c \phi) \cdot v \rangle + \langle 
 \sigma^\prime(\ell^c \phi \rightarrow \ell 
\phi^\dagger) \cdot v \rangle \right] \nonumber 
\end{eqnarray}
This is coupled with the differential equation for the density of the
Majorana neutrinos,
which reads:
\begin{equation}
\dot{n}_{\Psi_1} + 3 H n_{\Psi_1} = - \langle \Gamma_{\Psi_1} \rangle
    (n_{\Psi_1} - n_{\Psi_1}^{EQ}) - \frac{1}{2}
\left( \frac{n_L}{n_\gamma} \right) 
  n_{\Psi_1}^{EQ} (\varepsilon^\prime + \delta) 
 \langle \Gamma_{\Psi_1} \rangle \quad . \label{mama}
\end{equation}
The integration is now straight forward. We can integrate 
equation(\ref{mama}), then introduce the answer in equation (\ref{abi}) 
whose numerical integration gives the lepton asymmetry. The terms 
proportional to $n_L$ in equation (\ref{abi}) are multiplied by
negative coefficients and after integration, if they were alone, they
would wash out any lepton asymmetry exponentially.

\section{Numerical Integrations}
Following the usual method to evaluate the coupled Boltzmann equations
\cite{kolb} we introduce the dimensionless quantity 
\begin{equation}
z = \frac{M_1}{T} \quad ,
\end{equation}
to change the integration variable from the time $t$ to $z,$ which has
a temperature dependence. Furthermore it is convenient to look at the
particle density per comoving volume, which is:
\begin{equation}
Y = \frac{n}{s} \approx \frac{n}{g_\ast n_\gamma} \quad ,
\end{equation}
with $s$ being the entropy density and $g_\ast$ the number of 
effectively massless degrees of freedom (particles with mass $m \ll T$).
Now the differential equations (\ref{abi}) and (\ref{mama}) read:
\begin{eqnarray}
\frac{dY_L}{dz} & = & \frac{\langle \Gamma_{\Psi_1}(z=1) \rangle}{H(z=1)} z
\left\{ (\varepsilon^\prime + \delta)(Y_{\Psi_1} - Y_{\Psi_1}^{EQ})
\gamma - \frac{1}{2} Y_L \gamma_L \right\} \nonumber \\
\frac{dY_{\Psi_1}}{dz} & = & - \frac{\langle \Gamma_{\Psi_1}(z=1) 
\rangle}{H(z=1)} z \gamma \left\{ Y_{\Psi_1} - Y_{\Psi_1}^{EQ} (1 +
O(\varepsilon^\prime, \delta)) \right\} \label{horn}
\end{eqnarray}
with
\begin{displaymath}
Y_{\Psi_1}^{EQ} = \frac{1}{2g_\ast} \int_z^\infty dx \sqrt{x^2
-z^2} \frac{x}{e^x + 1} =
\left\{ \begin{array}{cl} 1/g_\ast & z \ll 1 \\ 
 1/g_\ast \sqrt{\pi/2} \, z^{3/2} \, e^{-z} & z \gg 1 \end{array}
\right.
\end{displaymath}
\begin{displaymath}
\gamma = \frac{\langle \Gamma_{\Psi_1}(z) \rangle}{
\langle \Gamma_{\Psi_1}(z=1) \rangle} = \frac{K_1(z)}{K_2(z)} 
= \left\{ \begin{array}{ll} z &
 z \ll 1 \\ 1 & z \gg 1 \end{array} \right. \nonumber
\end{displaymath} 
where $K_n$ are the modified Bessel functions \cite{wolf}, and
\begin{eqnarray}
\gamma_L(z) & = & \frac{
\left\{ \frac{1}{2} g_\ast Y_{\Psi_1}^{EQ} \langle \Gamma_{\Psi_1}(z) 
\rangle + 2 \langle \sigma^\prime v \rangle n_\gamma \right\}}{
\langle \Gamma_{\Psi_1}(z=1) \rangle} \nonumber \\
& \simeq & \left\{ \begin{array}{cl} z + 0.1 \, z^{-1} &
 z \ll 1 \\ z^{3/2} e^{-z} + 0.1 \, z^{-5} & z \gg 1 
\end{array} \right. \nonumber
\end{eqnarray}
Here we approximate the cross-section terms as in \cite{kolb}. For 
a detailed calculation of the cross-section terms see \cite{mich}.

The overall factor 
\begin{equation}
K = \frac{\langle \Gamma_{\Psi_1}(z=1) \rangle}{H(z=1)} 
\end{equation}
is an important measure for the efficiency to create a lepton
asymmetry. In the case $K \ll 1,$ which corresponds to $ 
\langle \Gamma_{\Psi_1}(z=1) \rangle \ll H(z=1),$ the decay rate
is much smaller than the expansion rate of the universe and the
particles come out of equilibrium and create a lepton asymmetry.
In the other case, where $K \gg 1,$ the particle decay fast and recombine
and are in thermal equilibrium in comparison to
the expansion rate of the universe. In this case the lepton asymmetry
will go to zero. 

Figure 2 shows this dependence of the lepton asymmetry, which we obtain
by integrating
the Boltzmann equations (\ref{horn}) numerically. The shape of
the curves is in direct correspondence to \cite{kolb}, 
but the asymptotic value
of the curves is now bigger, because of the $CP-$parameter $\delta$
coming out of the interference of the tree with the self energy diagram:
\begin{equation}
Y_L (z \rightarrow \infty) = \frac{\varepsilon^\prime + \delta}{g_\ast}
\quad , \quad {\rm for} \quad K \ll 1
\end{equation}
If $K$ increases the asymptotic value of the lepton asymmetry decreases and
finally for $K \gg 1$ goes to zero. Similar numerical results were
obtained in \cite{utpal} where the energy scale for the generation of
the lepton asymmetry was studied.

\begin{figure}[h]
\setlength{\unitlength}{1cm}
\begin{center}
\vspace{1cm}
\begin{picture}(13,8)
\epsfbox{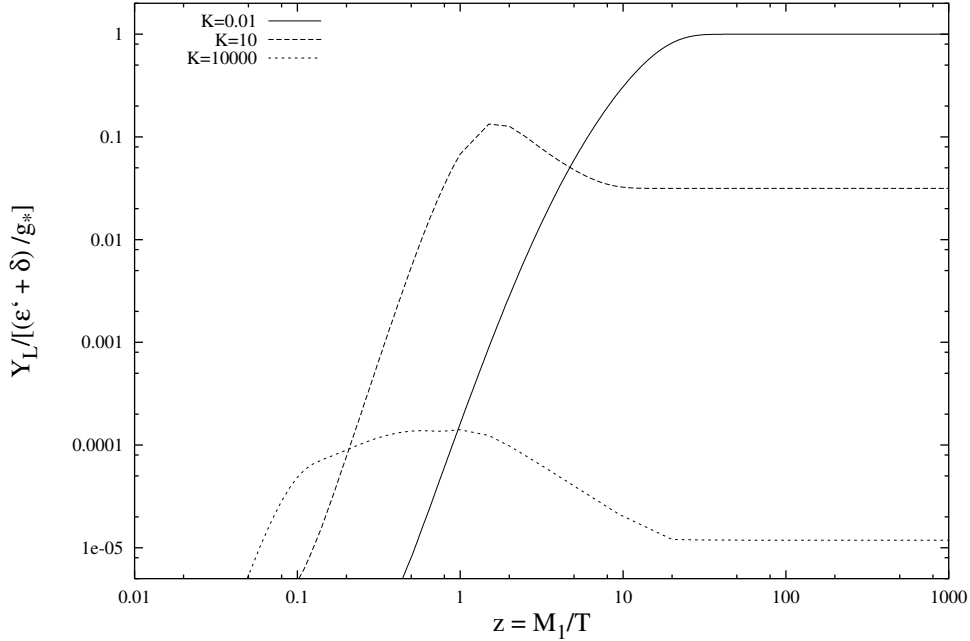}
\end{picture}
\vspace{-.3cm}
\caption{The development of the lepton asymmetry $Y_L / \left[
(\varepsilon^\prime + \delta)/g_\ast \right]$ with the expansion of
the universe given by $x = M_1/T$}
\end{center}
\end{figure}

\section{Conclusions}
We have shown explicitly how a lepton asymmetry is created
through the decay of Majorana neutrinos. The asymmetry is proportional to 
the sum of the self energy and the vertex contribution. 
We emphasize that at thermal equilibrium, the
unitarity of the amplitudes implies the vanishing of the lepton
asymmetry, as it is demonstrated in section 1. On the other hand,
the thermal development
of the Majorana densities is the reason for the unitarity cancellation to
become ineffective and allow the generation of the lepton asymmetry
(see section 2). 
Finally, we solve the Boltzmann equations numerically and
create a final asymmetry, which is proportional to the sum of the
$CP-$parameters coming from the vertex contribution $\varepsilon^\prime$
and from the self energy $\delta.$ The latter one has a resonant
behaviour depending on the mass difference of the Majorana particles
\cite{flanzjan}. 

\vspace{0.5cm}

%{\Large{\bf Acknowledgements}}
We wish to thank A. Pilaftsis for helpful communications on
unitarity and the incoherence of the physical states. Helpful discussions
with A.D. Dolgov and A. Joshipura are gratefully acknowledged. 

\noindent
\begin{appendix}
\section{The Mixed Majorana States}
The problem of mixing of Majorana neutrinos has been treated at
several places. One formalism is to discuss the development of
the various components in the wavefunction, where the absorptive
part has a definite effect \cite{flanz, flanzjan}. Another method
is to construct the inverse propagator and their renormalization
\cite{apost, apostel}. In both cases the absorptive part is a
physical observable, which can not be
removed by renormalization \cite{apostel}. We follow here the approach of
references \cite{flanz, flanzjan}, where to $O(h^2_{\alpha i})$ the
$CP-$effects were computed exactly for all relative values of the masses
\cite{flanzjan}.

We assume a mass hierarchy and consider an epoch
in the development of the universe, where the Majorana neutrinos
are decoupled from each other. When the temperature of the universe becomes
of the order of the lightest mass $M_1,$ there is only this particle left,
which can only decay bringing the universe out-of-equilibrium.  
The physical heavy Majorana Neutrino $\Psi_1$ is a mixed state of the
interaction states: 
\begin{equation}
|\Psi_1 \rangle = \frac{1}{\sqrt{2}} \left\{ |N_{R1} \rangle
  + |(N_{R1})^c \rangle + \alpha_2 |N_{R2} \rangle +
  \alpha_1 |(N_{R2})^c \rangle \right\} \label{trudy}
\end{equation}
It has a definite mass and equation (\ref{trudy}) gives the 
couplings of the various components. The transition amplitudes 
of the mixed state receive contributions from the
vertex corrections through the $\varepsilon^\prime$ and from
the self energies through the mixing of the states:
\begin{eqnarray}
|{\cal M}(\Psi_1 \rightarrow \ell_\alpha \phi^\dagger)|^2 & = &
|\langle \Psi_1 | N_{R1} \rangle \langle N_{R1} | \ell_\alpha 
 \phi^\dagger \rangle + \langle \Psi_1 | N_{R2} \rangle \langle 
 N_{R2} | \ell_\alpha \phi^\dagger \rangle|^2 \nonumber \\
& = & |{\cal M}_0|^2 \left\{ \frac{1}{2} (1 + \varepsilon^\prime) +
   \frac{2 \rm{Re} (\alpha_2 h_{\alpha_1}^\ast h_{\alpha 2})}
   {|h_{\alpha 1}|^2} + \, O \left( |\alpha_2|^2 \right) \right\} 
 \nonumber \\
 & = & |{\cal M}(\ell_\alpha^c \phi \rightarrow \Psi_1)|^2 
\end{eqnarray}
\begin{eqnarray}
|{\cal M}(\Psi_1 \rightarrow \ell_\alpha^c \phi)|^2 & = &
|\langle \Psi_1 | N_{R1}^c \rangle \langle N_{R1}^c | \ell_\alpha^c 
 \phi \rangle + \langle \Psi_1 | N_{R2}^c \rangle \langle 
 N_{R2}^c | \ell_\alpha^c \phi \rangle|^2 \nonumber \\
 & = & |{\cal M}_0|^2 \left\{ \frac{1}{2} (1 - \varepsilon^\prime) +
   \frac{2 \rm{Re} (\alpha_1^\ast h_{\alpha_1}^\ast h_{\alpha 2})}
   {|h_{\alpha 1}|^2} + \, O \left(|\alpha_1|^2 \right) \right\}
 \nonumber \\
 & = & |{\cal M}(\ell_\alpha \phi^\dagger \rightarrow \Psi_1)|^2
\end{eqnarray}
with $|{\cal M}_0|^2 = \sum_\alpha |h_{\alpha 1}|^2 M_1^2.$ These are 
the relevant amplitudes for the Boltzmann equation (\ref{bolle}), which
are given now by:
\begin{eqnarray}
\dot{n}_L + 3 H n_L & = & \Lambda_{12}^3 |{\cal M}_0|^2 \left[
- f_\ell f_{\phi^\dagger} \left\{ \frac{1}{2} (1 - \varepsilon^\prime)
+ \frac{\rm{Re} (\alpha_1^\ast h_{\alpha_1}^\ast h_{\alpha 2})} 
{|h_{\alpha 1}|^2} \right\} \right. \nonumber \\
 & & + f_{\ell^c} f_\phi \left\{ \frac{1}{2} (1 + \varepsilon^\prime)
+ \frac{\rm{Re} (\alpha_2 h_{\alpha_1}^\ast h_{\alpha 2})} 
{|h_{\alpha 1}|^2} \right\} \\
 & & + f_{\Psi_1} \left. \left\{ \varepsilon^\prime + 
\frac{\rm{Re} (h_{\alpha_1}^\ast h_{\alpha 2} (\alpha_2 - \alpha_1^\ast))} 
{|h_{\alpha 1}|^2} \right\} \right] \nonumber \\
 & & + 2 \, \Lambda_{12}^{34} \left\{ f_{\ell^c} f_\phi \left[
     |{\cal M}(\ell^c \phi \rightarrow \ell \phi^\dagger)|^2 -
     |{\cal M}_{RIS}(\ell^c \phi 
      \rightarrow \ell \phi^\dagger)|^2 \right] \right. \nonumber \\
  & & - \left. f_\ell f_{\phi^\dagger} \left[
     |{\cal M}(\ell \phi^\dagger \rightarrow \ell^c \phi)|^2 -
     |{\cal M}_{RIS}(\ell \phi^\dagger 
      \rightarrow \ell^c \phi)|^2 \right] \right\} \nonumber
\end{eqnarray}
Next we substitute the phase space densities and develop them
in the parameter $\mu/T:$
\begin{eqnarray}
f_\ell f_{\phi^\dagger} & = & f_{\Psi_1}^{EQ} e^{\mu/T} = f_{\Psi_1}^{EQ}
    \left[ 1 + \frac{\mu}{T} + O\left( \frac{\mu^2}{T^2} \right) \right] 
 \nonumber \\
f_{\ell^c} f_{\phi} & = & f_{\Psi_1}^{EQ} e^{-\mu/T} = f_{\Psi_1}^{EQ}
    \left[ 1 - \frac{\mu}{T} + O\left( \frac{\mu^2}{T^2} \right) \right] 
\end{eqnarray}
As usual we assume that the particles are in kinetic equilibrium.  
Finally, we use the definition of the thermally-averaged decay width
$\langle \Gamma \rangle$ which is:
\begin{equation}
\langle \Gamma \rangle = \frac{g/(2\pi)^3 \int d^3p f(p) \Gamma(p)}
 {g/(2\pi)^3 \int d^3p f(p)} = \frac{g/(2\pi)^3 \int d^3p f(p) \Gamma(p)}
 {n}
\end{equation}
where $g$ are the degrees of freedom and $n$ is the particle density. 
With all this information we arrive at equation(\ref{wild}) and
equation (\ref{firle}). In the next section we study the scattering
terms in the Boltzmann equation.

\section{The Scattering Amplitudes}
In this section we study the scattering terms in the Boltzmann equation:
\begin{eqnarray}
-2 \Lambda_{12}^{34} f_{\Psi_1}^{EQ} \left\{ \left[ 
|{\cal M}_{RIS}(\ell^c \phi \rightarrow \ell \phi^\dagger)|^2 -
|{\cal M}_{RIS}(\ell \phi^\dagger \rightarrow \ell^c \phi)|^2 \right]
\right. \nonumber \\
\left. + \frac{\mu}{T} \left[ |{\cal M}^\prime(\ell \phi^\dagger 
\rightarrow \ell^c \phi)|^2 + |{\cal M}^\prime(\ell^c \phi \rightarrow 
\ell \phi^\dagger)|^2 \right] \right\} \label{fifi}
\end{eqnarray} 
with
\begin{equation}
|{\cal M}^\prime|^2 = |{\cal M}|^2 - |{\cal M}_{RIS}|^2
\end{equation}
For the real intermediate states one can use the narrow-width 
approximation:
\begin{eqnarray}
|{\cal M}_{RIS}(\ell \phi^\dagger \rightarrow \ell^c \phi)|^2
& = & \frac{\pi}{M_1 \Gamma_{\Psi_1}} \delta(s-M_1^2)
|{\cal M}(\ell \phi^\dagger \rightarrow \Psi_1)|^2 
|{\cal M}(\Psi_1 \rightarrow \ell^c \phi)|^2 \nonumber \\ \\
|{\cal M}_{RIS}(\ell^c \phi \rightarrow \ell \phi^\dagger)|^2
& = & \frac{\pi}{M_1 \Gamma_{\Psi_1}} \delta(s-M_1^2)
|{\cal M}(\ell^c \phi \rightarrow \Psi_1)|^2 
|{\cal M}(\Psi_1 \rightarrow \ell \phi^\dagger)|^2 \nonumber
\end{eqnarray}
Using the following integral \cite{wolf}:
\begin{equation}
\int d\Pi_1 d\Pi_2 f_{\Psi_1}^{EQ} \delta(s-M_1^2) = \frac{1}{(2\pi)^6}
2 \pi^4 \frac{n_{\Psi_1}^{EQ}}{M_1 \Gamma_{\Psi_1}} \langle \Gamma_{\Psi_1}
\rangle
\end{equation}
and the definition of the thermally-averaged cross section
\begin{eqnarray}
\langle \sigma \cdot v \rangle & = & \frac{g_A/(2\pi)^3 g_B/(2\pi)^3
\int d^3p_A \int d^3p_B f(\vec{p}_A) f(\vec{p}_B) \sigma(AB 
\rightarrow CD) v}{g_A/(2\pi)^3 g_B/(2\pi)^3
\int d^3p_A \int d^3p_B f(\vec{p}_A) f(\vec{p}_B)} \nonumber \\ \\
& = & \frac{g_A/(2\pi)^3 g_B/(2\pi)^3
\int d^3p_A \int d^3p_B f(\vec{p}_A) f(\vec{p}_B) \sigma(AB 
\rightarrow CD) v}{n_A n_B} \nonumber
\end{eqnarray}
as well as the approximations $n_\ell \approx n_{\ell^c} \approx n_\phi
\approx n_{\phi^\dagger} \approx n_\gamma,$ expression (\ref{kai})
can be derived from equation (\ref{fifi}).

\end{appendix}

\end{document}